\documentclass[12pt]{article}
\usepackage[dvipdfmx]{graphicx}

\usepackage{amssymb}
\usepackage{url}
\usepackage{array}
\usepackage{amsmath}
\usepackage{latexsym}
\usepackage{setspace} 
\usepackage{comment}
\usepackage{subfigure}
\usepackage{slashbox}

\begin{document}

\begin{titlepage}
  \begin{center}

    \vspace{0.4cm}
           {\large\bf Study of top quark pair production near threshold at the ILC}
           \vspace{0.8cm}
           
           {\bf Tomohiro Horiguchi}$^{(a)}$, {\bf Akimasa Ishikawa}$^{(a)}$, {\bf Taikan Suehara}$^{(a)}$, 
           \\{\bf Keisuke Fujii}$^{(b)}$, {\bf Yukinari Sumino}$^{(a)}$, {\bf Yuichiro Kiyo}$^{(c)}$, 
           \\ and {\bf Hitoshi Yamamoto}$^{(a)}$
           
           \vspace{0.8cm}
           $^{(a)}Department \ of \ Physics,\  Tohoku \ University,\  Sendai, \ Miyagi, \ Japan$ \\
           $^{(b)}High \ Energy \ Accelerator \ Research \ Organization \ (KEK), \ Tsukuba, \ Ibaraki, \ Japan$ \\
           $^{(c)}Department \ of \ Physics, \ Juntendo \ University, \ Inzai, \ Chiba, \ Japan $\\
           \vspace{0.8cm}
           
           \normalsize{{\it Abstract} } \\
           
  \end{center}           
  
  We report on a study of top pair production at the International Linear Collider (ILC) 
  around center of mass energy (E$_{\rm CM}$) = 350 GeV using an ILD detector simulator based on 
  the Detailed Baseline Design (DBD) configuration. 
  Here we will report on a result of 6-Jet final state, $t\overline{t} \rightarrow bWbW \rightarrow bqqbqq$. 
  A result for the 4-Jet final state, $t\overline{t} \rightarrow bWbW \rightarrow bqqbl\nu$, 
  which has almost the same statics as that of the 6-Jet final state will be included in the future.
  For an energy scan of 11 center of mass energy points (340 - 350GeV) and two beam polarization 
  combinations (P($e^+,\ e^-$) = ($\pm$0.3, $\mp$0.8)) with 10 fb$^{-1}$ each, the statistical errors 
  on the top quark Yukawa coupling, its mass and width are estimated. 
  The results are $\delta y_t$ = 4.2\%, $\delta m_t$ = 16 MeV in potential subtracted scheme (PS), 
  and $\delta \Gamma_t$ = 21 MeV.

\end{titlepage}

\setcounter{footnote}{0}

\section{Introduction}
\label{sec:intro}

The top quark is the heaviest particle in the Standard Model~(SM). 
Since the top quark mass measured at hadron colliders, $m_t = 173.1 \pm 0.9$~GeV\cite{PDG13}, 
is close to the electroweak scale, $v/\sqrt{2} = 174$~GeV, top quark may play an important role 
in the electroweak symmetry breaking. 
However, top quark mass measured at the hadron colliders is a Monte Carlo parameter,
which is very hard to translate into masses defined in other schemes used for theoretical calculation, 
such as the $\overline{\rm MS}$ scheme\cite{MSbar}. 
With recent measurements of the Higgs boson mass at the LHC and the top quark $\overline{\rm MS}$ mass, 
$m_t^{\overline{\rm MS}} = 160 \pm 5$~GeV\cite{D0msbar}, derived from the top pair 
production cross section, the vacuum stability can be discussed in the Standard Model.
Although the uncertainties are still large, the measurements suggest that the vacuum 
of our universe might be meta-stable\cite{vaccumestability}. 
To draw a definite conclusion, a precise measurement of the top quark mass in a 
theoretically calculable scheme is essential.     

Since the top quark decays very quickly due to its heaviness, the width of the top quark is sizable, 
about 1.4~GeV, which is predicted in the next-to-next-to-leading order (NNLO) calculation in the SM. 
The top quark width is an important probe for anomalous couplings and exotic decays. Since the experimental 
resolution for the top quark mass at the hadron colliders is much worse than the top quark width, 
its direct measurement is impossible. The D0 Collaboration indirectly measured the top quark width from 
the t-channel single top production cross section and the branching fraction of the top decaying into a bottom 
quark and a $W$ boson\cite{D0width}. The result, $2.1 \pm 0.6$~GeV, is consistent with the SM prediction though 
the error is still large to discuss exotic contributions.

The top quark Yukawa coupling ($y_t$) is a fundamental parameter in the SM, and also important 
for physics beyond the SM (BSM), since the top quark Yukawa coupling enters renormalization 
group equations for many BSM parameters. Recent measurements of Higgs boson production 
in gluon-gluon fusion at the LHC give an indirect constraint on the top quark Yukawa coupling. 
In the future, a measurement of Higgs production off top or anti-top quarks, $t\bar{t}H$, at the LHC will 
directly constrain the top quark Yukawa coupling but its precise determination is difficult 
due to reconstruction under large QCD background and large theoretical uncertainty.

\begin{figure}[tphb]
  \begin{center}
    \includegraphics[width=6cm]{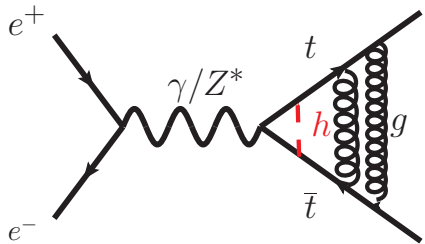} 
    \caption{Higgs exchange diagram for top pair production}
    \label{fig:higgsexchange}
  \end{center}
\end{figure}

The ILC is an ideal place to measure top quark properties due to the following reasons.
Firstly, since it only involves interaction of elementary particles (electrons and positrons), 
the center of mass energy (E$_{\rm CM}$) is tunable and its spread is reasonably small\cite{DBDbeamparam}. 
Secondly, the detector exploits state-of-the-art technology, allowing high precision measurements in 
triggerless operation with no inefficiency in data acquisition thanks to low beam-beam backgrounds.
Lastly, uncertainties in theoretical calculations are much lower than at the hadron colliders.
Thus one can perform a precision top pair threshold scan at E$_{\rm CM}$ around 350~GeV and extract 
the top quark mass and width in a theoretical clean manner. 
The top quark Yukawa coupling can also be derived from precise measurements of the cross sections 
since it is enhanced by Higgs exchange diagrams (Fig.\ref{fig:higgsexchange}),  which is proportional to $y_t^2$.

The measurement accuracies of top quark mass, width, and yukawa coupling have been studied 
by the previous work\cite{jlc}\cite{Martinez:2002st} and at the CLIC\cite{clic}. We report the latest result at the ILC.

This paper is organized as follows. In Section \ref{sec:framework}, the framework used for 
the analysis is described. Section \ref{sec:eventselection} explains our event reconstruction 
and selection designed to suppress backgrounds. Section \ref{sec:analysis} is devoted to the extractions of 
the top quark Yukawa coupling, its mass and width with estimations of the statistical errors on these parameters. 
Finally, in Section \ref{sec:summary}, we summarize the results.

\section{Analysis Framework}\label{sec:framework}

\subsection{Signal and Backgrounds}
Since a top quark decays into a bottom quark and a $W$ boson, the decay channels are categorized by the decay of 
the two $W$ bosons. If the final states of the two $W$s are 4 quarks, 2 quarks + 1 lepton + 1 neutrino, 
or 2 leptons + 2 neutrinos, the channels are hereafter called 6-Jet, 4-Jet and 2-Jet, respectively. 
In this paper, a 6-Jet analysis will be reported and the other top pair channels are treated as backgrounds. \\
Processes with high jet multiplicity and a comparable cross section to that of the top pair production 
such as $WW \rightarrow qqqq$, $ZZ \rightarrow qqqq$, and $ZH$($ZH \rightarrow qqH$) are considered as backgrounds. 
Other backgrounds with 6-Jet final states such as $tbW$, $WWZ$, and $ZZZ$ will be added in the future.   

\subsection{Polarization and Integrated luminosity}
The running scenario around the ILC E$_{\rm CM}$ = 350 GeV, an operation as the top factory, has not been decided yet. 
Here we assume the following settings. The data will be acquired using two polarization configurations
\begin{center} P($e^{+}$, $e^{-}$) = (+30\%, -80\%)  \end{center}
\begin{center} P($e^{+}$, $e^{-}$) = (-30\%, +80\%)  \end{center}
so that we can also separate top couplings to photon and $Z$ boson. Throughout this paper, the former (latter) 
polarization configuration is denoted as ``Left'' (``Right''). The top threshold scan is performed 
at 11 energy points, every 1 GeV from 340GeV to 350GeV for both polarization combinations 
with an integrated luminosity of 10 fb$^{-1}$ each, which amount to 220 fb$^{-1}$ in total. 

\subsection{Event Generators}

To estimate the signal efficiencies and background yields, event samples at around E$_{\rm CM}$ = 350 GeV 
were generated by Monte-Carlo event generators. For signal top pairs, Physsim\cite{physsim} was used as 
an event generator, while for backgrounds, Whizard\cite{whizard} was utilized. Both of them are of 
Leading Order (LO). The top quark pole mass was set at 174~GeV/c$^2$. Physsim is based 
on full helicity amplitudes including QCD enhancement near the top pair threshold,
calculated using HELAS\cite{helas}, which properly takes into account the angular correlations of 
the decay products. Parton-showering and fragmentation of colored quarks and gluons are done 
by PYTHIA\cite{pythia} with parameter tuned by the OPAL collaboration.  
The beam parameters, which include initial state radiation, beamstrahlung, and beam energy spread, 
are common for both generators, as specified by a so called Lumi-linker file\cite{lumilinker}. 

\subsection{Detector Simulation}

The ILD detector has a tracker system and a finely segmented calorimeter 
system to realize the particle flow algorithm~(PFA)~\cite{TDR_detector}.
In addition, a solenoid magnet which provides 3.5~T magnetic field,
a muon tracker combined with iron yoke magnetic flux return,
and luminosity and beam monitors are equipped.

The tracker consists of three doublet layers of silicon pixel detector~(VTX),
a silicon strip tracker system, and a time projection chamber~(TPC).
The inner-most layer of the VTX is placed at 1.6~cm from the interaction point in radial direction
to achieve the impact parameter resolution of $\sigma_b<$~5 $\oplus$ 15/p sin$^{3/2}\theta$, 
which is necessary for excellent flavor tagging of heavy quarks, such as bottom and charm quarks.
The silicon strip tracker system consists of the silicon inner tracker~(SIT), 
the silicon external tracker~(SET), and the endcap tracking detector~(ETD).
The SIT is located in between the VTX and the TPC at barrel region. 
The SET and the ETD envelop the TPC at barrel and endcap regions, respectively.
The role of silicon strip tracker system is to improve the momentum resolution 
and to extrapolate charged tracks from the TPC to the VTX or to the calorimeter system.
The TPC is a gaseous cylindrical chamber with inner and outer radii of 329~mm
and 1808~mm and half-length of $\pm$ 2350~mm in $Z$ direction.
The spacial resolution of TPC in the r-$\phi$ direction is better than 100$\mu$m and
two-track separability is about 2~mm.
The momentum resolution using only TPC is $\sigma(1/p) \sim 10^{-4}$~GeV$^{-1}$,
and this can be improved to $\sigma$(1/p)~$\approx$~2~$\times$~10$^{-5}$~GeV$^{-1}$
with a full tracking system.

The calorimeter system comprises an electromagnetic calorimeter~(ECAL),
which measures energy of photons and identifies electrons,
and a hadronic calorimeter~(HCAL), which measures energy of neutral hadrons.
Both the ECAL and HCAL are finely segmented sampling calorimeters
to detect showers from individual particles 
in order to give the excellent performance with the PFA.
The absorber and active material for the ECAL and HCAL are tungsten and silicon pixel sensors,
iron and tile scintillators with MPPCs, respectively.

We used the MOKKA\cite{Mokka} detector simulation tool based on GEANT4 to describe the ILD detector.
With the ILD detector simulation and a particle flow algorithm called PandraPFA~\cite{pandora}, 
a dijet energy resolution of 25~\%/$\sqrt{\rm E_{jj}(GeV)}$ for 45~GeV dijets can be achieved.

\section{Event Reconstruction and Selection} \label{sec:eventselection}
\subsection{Event Reconstruction}
After the detector simulation, PFOs (Particle Flow Objects) were clustered to jets using Durham algorithm. 
In the Durham algorithm, each PFO is regarded as a jet on its own to begin with, a jet pair $i$ and $j$ 
gets combined if the pair has the lowest $Y_{ij}$ value which is defined as
\begin{equation}
  Y_{ij} =  \cfrac{2\min\{E_i^2, E_j^2\}(1-\cos\theta_{ij})}{E_{vis}^2}     \nonumber 
\end{equation} 
where $\theta_{ij}$ is the angle between the momentum vectors of the two particles.
In this analysis, PFOs were forced to cluster into 6 jets. Among the 6 jets, two most $b-$like jets were identified 
using the LCFIPlus flavor tagging algorithm, using vertex and mass information. The two $W$s were reconstructed 
from the remaining 4 jets. The two top quarks were then reconstructed by pairing $b-$like jets with $W$ candidates. 
Since there were multiple possible ways to combine the jets, we defined a quantity called $\chi^2$ as
\begin{equation}
  \chi^2 = \cfrac{(m_{2j}-m_{W})^2}{\sigma_W^2} + \cfrac{(m_{2j}-m_{W})^2}{\sigma_W^2} 
  + \cfrac{(m_{3j}-m_{t})^2}{\sigma_t^2}+ \cfrac{(m_{3j}-m_{t})^2}{\sigma_t^2},  \nonumber
\end{equation} 
and chose the jet combination which minimized the $\chi^2$ value. Here $m_{2j}$($m_{3j}$) is 
the invariant mass from 2(3) jets. For the two jets used for m$_{2j}$, we do not use $b-$like jets. 
$m_t$ and $m_W$ are the top mass (174~GeV) and the $W$ mass~(80.0~GeV) defined in the generators, 
$\sigma_t$ and $\sigma_W$ are mass resolutions for the top quark and the $W$ boson.

\subsection{Event Selection}
After the event reconstruction, backgrounds from $t\overline{t}$ $\rightarrow$ 4-Jet and 2-Jet, $WW$, $ZZ$, 
and $ZH$ events were suppressed by various selection cuts to maximize the significance. 
Since the signal and background cross sections are different for ``Left'' and ``Right'' polarizations, 
we adopted different selection cuts for them.

\begin{figure}[tphb]
  \begin{center}
    \subfigure[Signal efficiency]{\includegraphics[width=7cm]{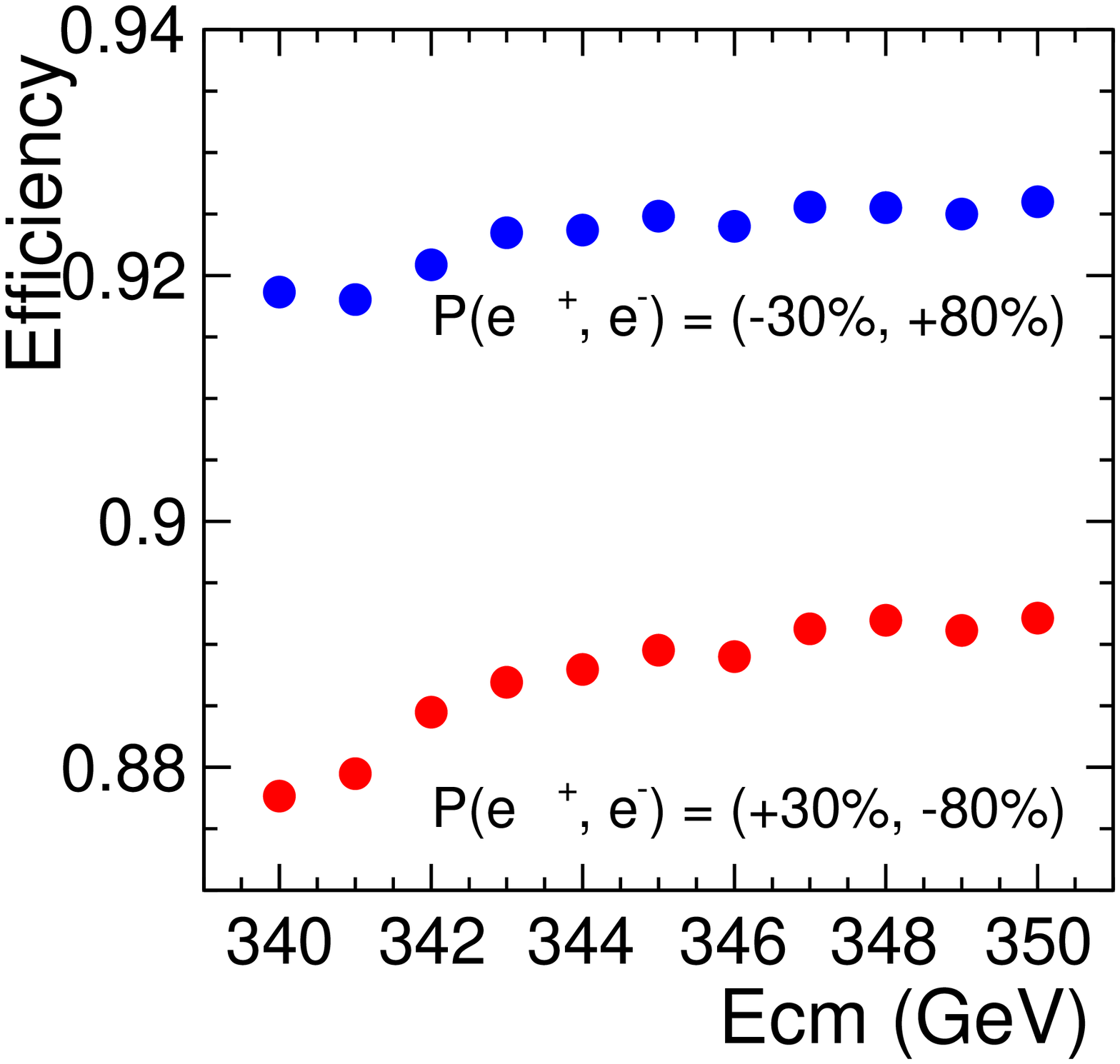} \label{fig:thcr1}}
    \subfigure[Cross section]{\includegraphics[width=7cm]{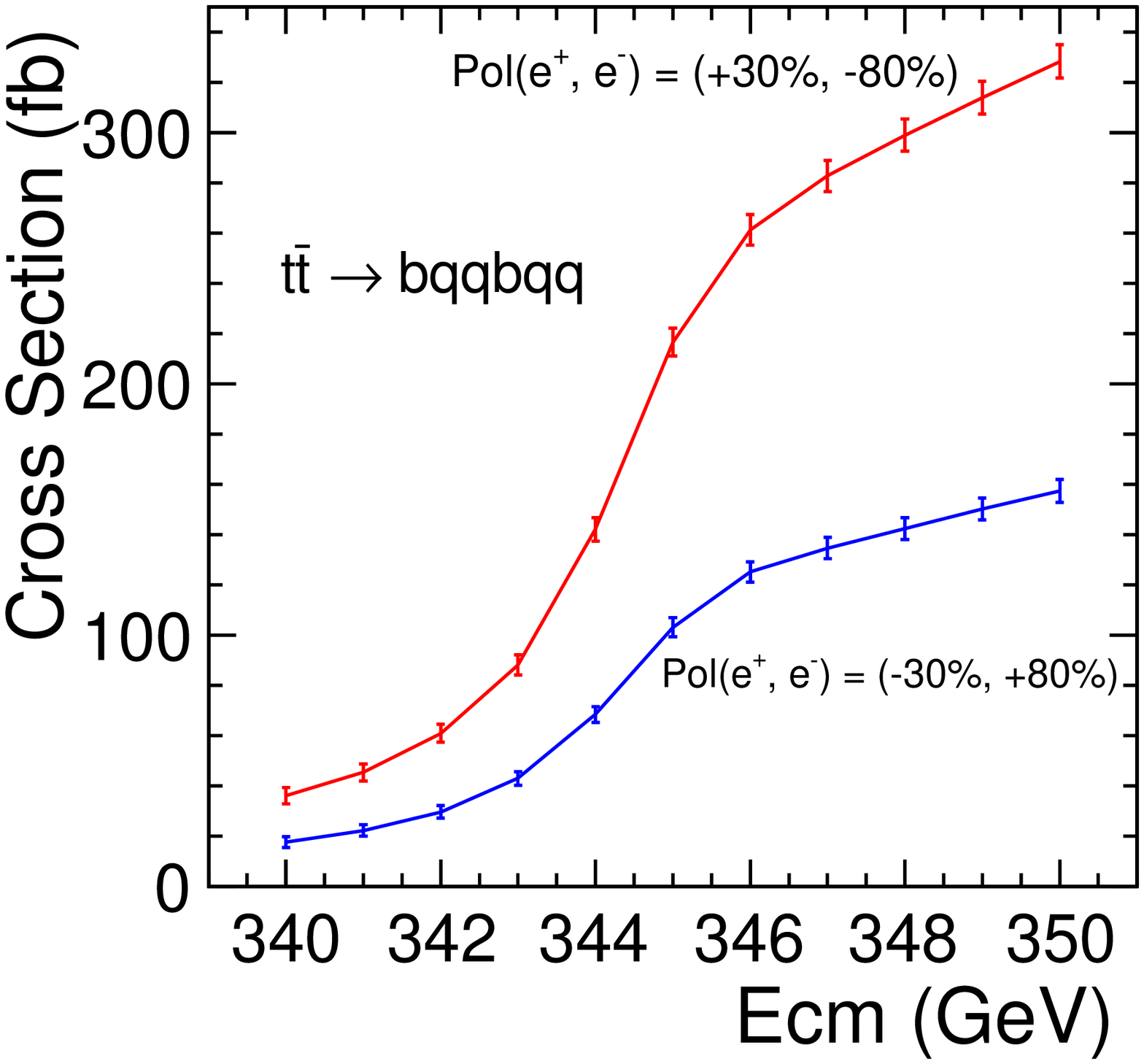} \label{fig:thcr2}}
    \caption{Signal selection efficiency and cross section (LO generator)}
    \label{fig:CrossEff}
  \end{center}
\end{figure}

The $WW$ background can be suppressed by requiring two $b-$like jets, because $W$ boson decay to a bottom quark is CKM 
suppressed while the branching fraction for top quark decay to a bottom quark and a $W$ boson is more than 99\%. 
After requiring two $b-$like jets, we applied cuts on thrust and visible energy. Isolated leptons were then looked for 
by using energy flow in a cone around a lepton candidate track. When the energy flow in this cone was lower than 5 GeV, 
we regarded this track as an isolated lepton. When some isolated lepton candidate was found, the event was discarded. 
The remaining events were subject to further cuts on the Durham $Y$ value ($Y_{45}$), missing transverse momentum 
($p_t^{\rm miss}$), and the number of PFOs (nPFOs). The numbers of signal and background events after 
each selection cut for the ``Left'' and ``Right'' polarizations are summarized in Table \ref{tb:cutL} and 
Table \ref{tb:cutR}. Notice that we generated the signal events for each energy point 
to estimate signal efficiency, while the background events were generated only at E$_{\rm CM}$ = 350 GeV. 
The background at other energy points were estimated by scaling the cross section to save CPU time. 
Fig \ref{fig:CrossEff} shows the obtained cross section and signal efficiency at each energy point.

\begin{table}[tphb]
  \begin{center}
    \begin{tabular}{|l|cccccc|c|} \hline
      E$_{\rm CM}$= 350(GeV) on ``Left''  & $t\overline{t}$ 6-Jet & $t\overline{t}$ 4-Jet & $t\overline{t}$ 2-Jet & $WW$  & $ZZ$ & $ZH$ & $S_{6\rm{-Jet}}$ \\ \hline
      Generated                      & 3288  & 3167  & 763   & 65328 & 6008 & 1389 & 11.6  \\ \hline
      btag1 \textgreater 0.1, btag2 \textgreater 0.1 & 3136 & 3004 & 725 & 7567 & 2832 & 982 & 23.2 \\ \hline
      thrust \textless 0.84          & 3090  & 2882  & 645   &  867  & 917  & 815  & 32.2  \\ \hline 
      Visible Energy \textgreater 310(GeV)  & 3063  & 1194  & 37    &  434  & 573  & 577  & 39.9 \\ \hline
      nlep = 0                       & 3021  & 399   & 3     &  429  & 571  & 571  & 42.8 \\ \hline
      $Y_{\rm 45}$ \textgreater 0.0012 , $Y_{\rm 56}$ \textgreater 0.0007 & 2956 & 331 & 2 & 174 & 176 & 193 & 47.8 \\ \hline
      p$_{\rm t}^{\rm miss}$ \textgreater 38(GeV)    & 2942  & 160   & 0     &  173  & 175  & 192  & 48.7  \\ \hline
      nPFOs = 95                     & 2917  & 137   & 0     &  115  & 143  & 170  & 49.4  \\ \hline
    \end{tabular}   
    \caption{The numbers of signal and background events, and significance ($S_{6\rm{-Jet}}$) after each selection cut for the center of mass energy of 350~GeV and the ``Left'' polarization with 10 fb$^{-1}$.}
    \label{tb:cutL}
  \end{center}
  
  \begin{center}
    \begin{tabular}[bt]{|l|cccccc|c|} \hline
      E$_{\rm CM}$= 350(GeV) on ``Right''  & $t\overline{t}$ 6-Jet & $t\overline{t}$ 4-Jet & $t\overline{t}$ 2-Jet & $WW$  & $ZZ$ & $ZH$ & $S_{6\rm{-Jet}}$ \\ \hline
      Generated                       & 1572  & 1515  & 365   & 4326  & 2773 & 937  & 14.7  \\ \hline
      btag1 \textgreater 0.065 , btag2 \textgreater 0.065   & 1546  & 1483 & 355  & 1181 & 1591 & 720 & 18.7  \\ \hline
      thrust \textless 0.84           & 1522  & 1425  & 318   & 141   & 424  & 594  & 22.9  \\ \hline 
      Visible Energy \textgreater 305(GeV)   & 1514  & 687   & 24    & 73    & 267  & 438  & 27.6  \\ \hline
      nlep = 0                        & 1495  & 224   & 2     & 72    & 265  & 431  & 29.9  \\ \hline
      $Y_{\rm 45}$ \textgreater 0.0014 , $Y_{\rm 56}$ \textgreater 0.0006 & 1472 & 189 & 1 & 30 & 89 & 161 & 33.4  \\ \hline
      p$_{\rm t}^{\rm miss}$ \textgreater 38(GeV)     & 1465  & 89    & 0     & 30    & 88   & 160  & 34.2  \\ \hline
      nPFOs = 95                     & 1453  & 74    & 0     & 18    & 66   & 140  & 34.7  \\ \hline
    \end{tabular}
    \caption{The numbers of signal and background events, and significance ($S_{6\rm{-Jet}}$) after each selection cut for the center of mass energy of 350~GeV and the ``Right'' polarization with 10 fb$^{-1}$.}
    \label{tb:cutR}
  \end{center}
\end{table}

\newpage
 
\section{Estimation of statistical errors on the top quark Yukawa coupling, its mass and width} \label{sec:analysis}

\subsection{Top quark Yukawa Coupling}  
Since the enhancement of the top quark pair production cross section due to Higgs exchange diagram is 
energy-independent, approximately 9\%, in the threshold region, we can combine the numbers of signal and background 
events at all the 11 energy points (340~GeV $\sim$ 350~GeV) to estimate statistical error on the top quark Yukawa 
coupling. The $t\overline{t}$ cross section can be expressed by amplitudes with and without higgs exchange as follows. 
\begin{eqnarray}
  \sigma_{t\overline{t}} \propto |{\cal M}_{t\overline{t}}|^2 
  &=& | {\cal M}_{no \ higgs \ exchange} + y_t^2{\cal M}_{higgs \ exchange}|^2  \nonumber  \\
  &\sim& | {\cal M}_{no \ higgs \ exchange}|^2 + 2y_t^2|{\cal M}_{no \ higgs \ exchange} \times 
        {\cal M}_{higgs \ exchange}|.             \nonumber
\end{eqnarray}
Since the exchanged Higgs boson couples to the top quark twice, the leading correction term due to Higgs exchange is 
proportional to $y_t^2$, which corresponds to approximately a 9\% enhancement. The $\mathcal{O}(y_t^4)$ term is 
small enough to ignore. The sensitivities to the top quark Yukawa coupling were estimated using the following formula.
\begin{eqnarray}
  \frac{\delta y_t}{y_t} &\sim& \cfrac{(100+9) \times \cfrac{1}{2} \times \cfrac{\delta \sigma}{\sigma}}{9}   \nonumber 
  \label{eq:yukawa}
\end{eqnarray}
The expected statistical errors on top quark Yukawa coupling are 5.0\% and 7.1\% for the ``Left'' and ``Right'' 
polarization combinations, 
respectively, and 4.2\% when combined (Table~\ref{tb:topyukawa}).

\newlength{\myheight} 
\setlength{\myheight}{10mm}
\begin{table}[tphb]
  \begin{center}
    \begin{tabular}[bt]{|c|c|c|c|} \hline
      \parbox[c][\myheight][c]{0cm}{} \slashbox[3.5cm]{}{}  & ``Left'' & ``Right'' & Combined \rule[0mm]{0mm}{5mm} \\ \hline
      \parbox[c][\myheight][c]{0cm}{} cross section       & 0.84 \% & 1.2 \% & \slashbox[3cm]{}{} \rule[0mm]{0mm}{5mm} \\ \hline
      \parbox[c][\myheight][c]{0cm}{} top quark Yukawa coupling & 5.0 \%  & 7.1 \% & 4.2 \%   \\ \hline 
    \end{tabular}   
    \caption{Expected statistical errors on the top pair production cross section and the top quark Yukawa coupling 
      for the ``Left'', the ``Right'' and combined polarization combinations}
    \label{tb:topyukawa}
  \end{center}
\end{table}

\subsection{Top mass and width}

The top mass and width can be determined with unprecedented sensitivity by performing threshold scan at the ILC.
We fit the two parameters simultaneously using the cross section values measured at the 11 energy points 
in the threshold region of 340 to 350 GeV.

\subsubsection {Mass scheme and assumptions}

\begin{figure}[tphb]
  \begin{center}
    \subfigure[Theoretical Cross section (fixed top width)]{\includegraphics[width=7cm]{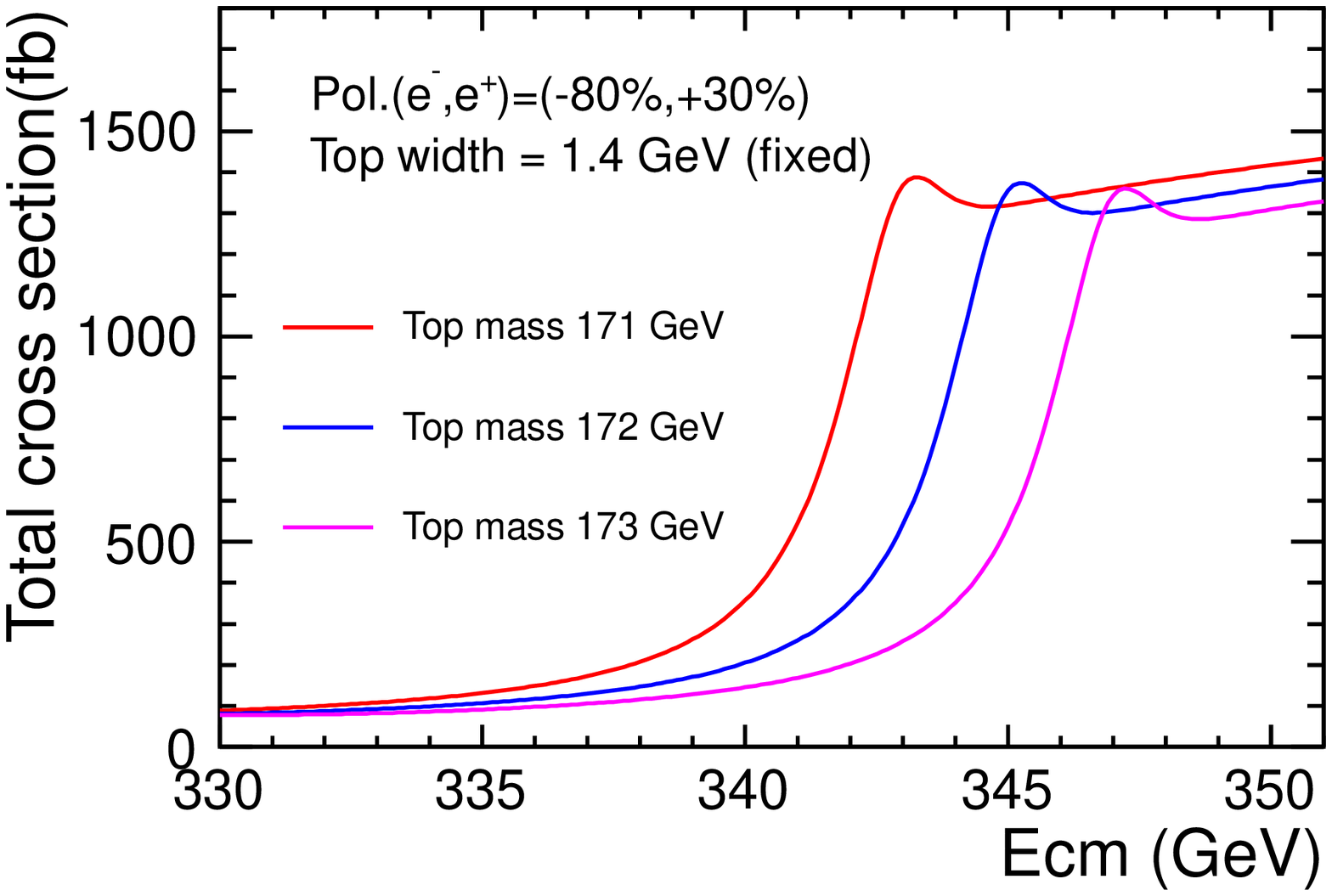} \label{fig:thcross1}}
    \subfigure[Theoretical Cross section (fixed top mass)]{\includegraphics[width=7cm]{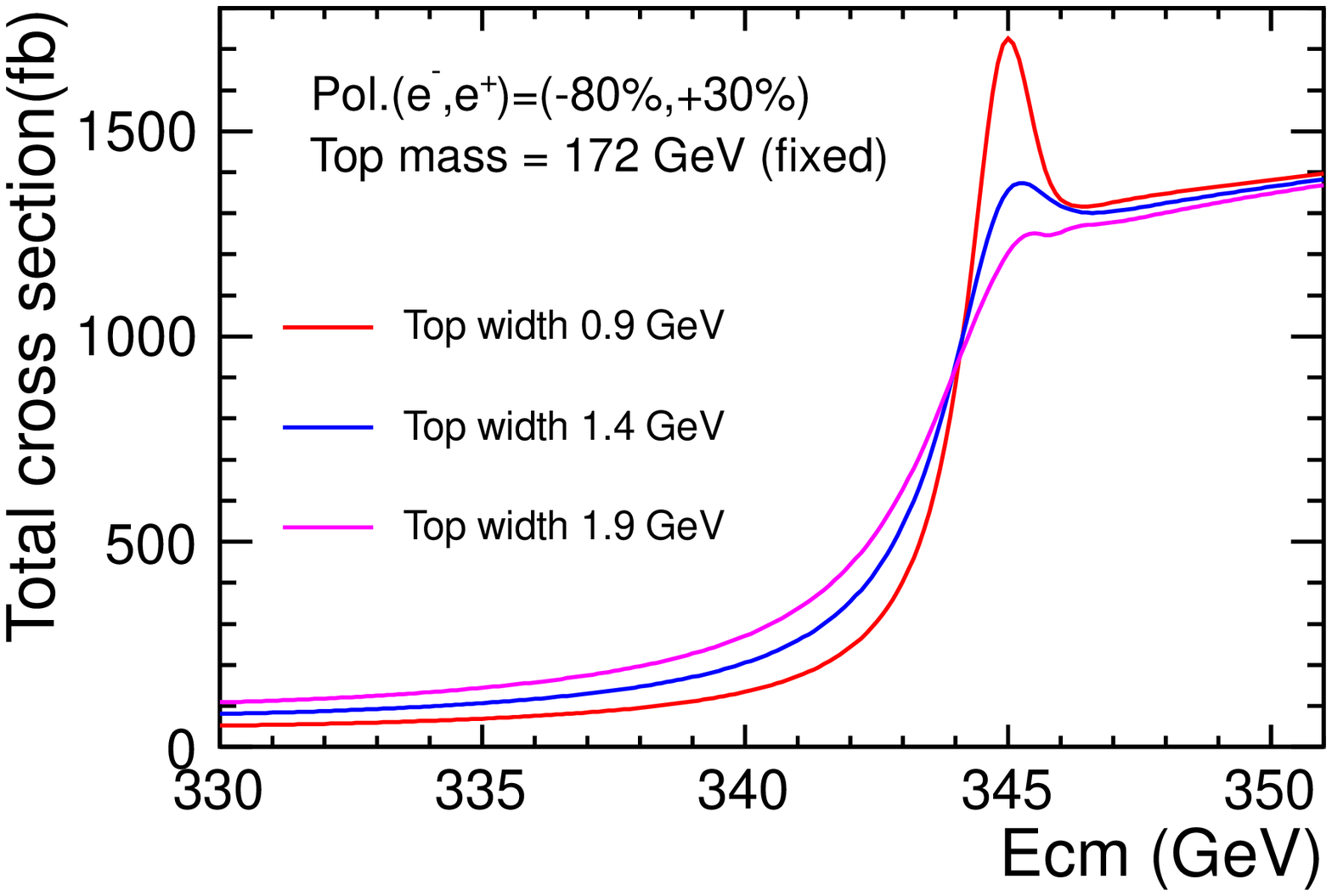} \label{fig:thcross2}}
    \caption{Theoretical Cross Section near top pair threshold}
    \label{fig:theocross}

    \subfigure[Luminosity Spectrum]{\includegraphics[width=7cm]{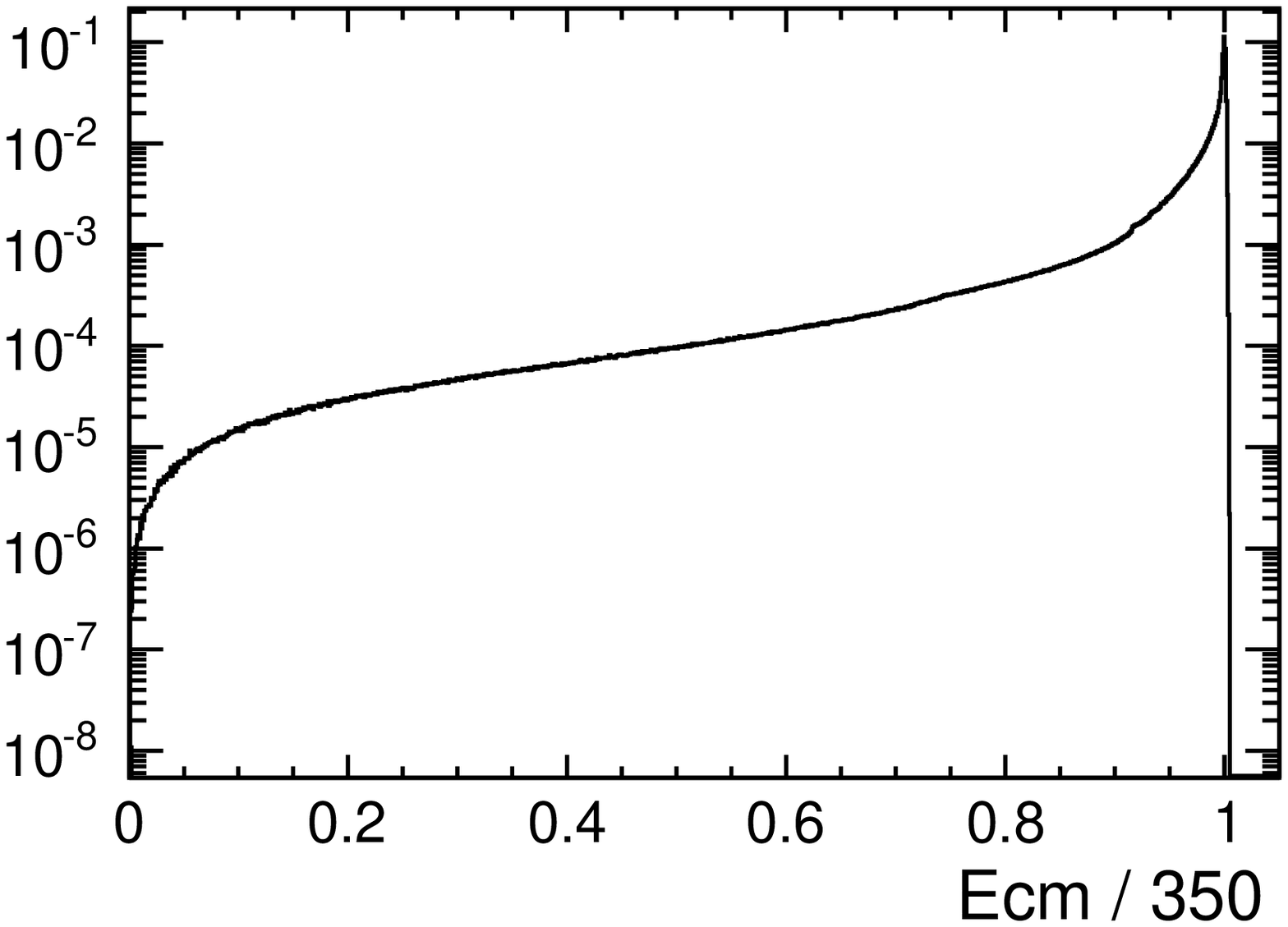} \label{fig:lumispe}}
    \subfigure[Luminosity Spectrum (blow-up of the peak region )]{\includegraphics[width=7cm]{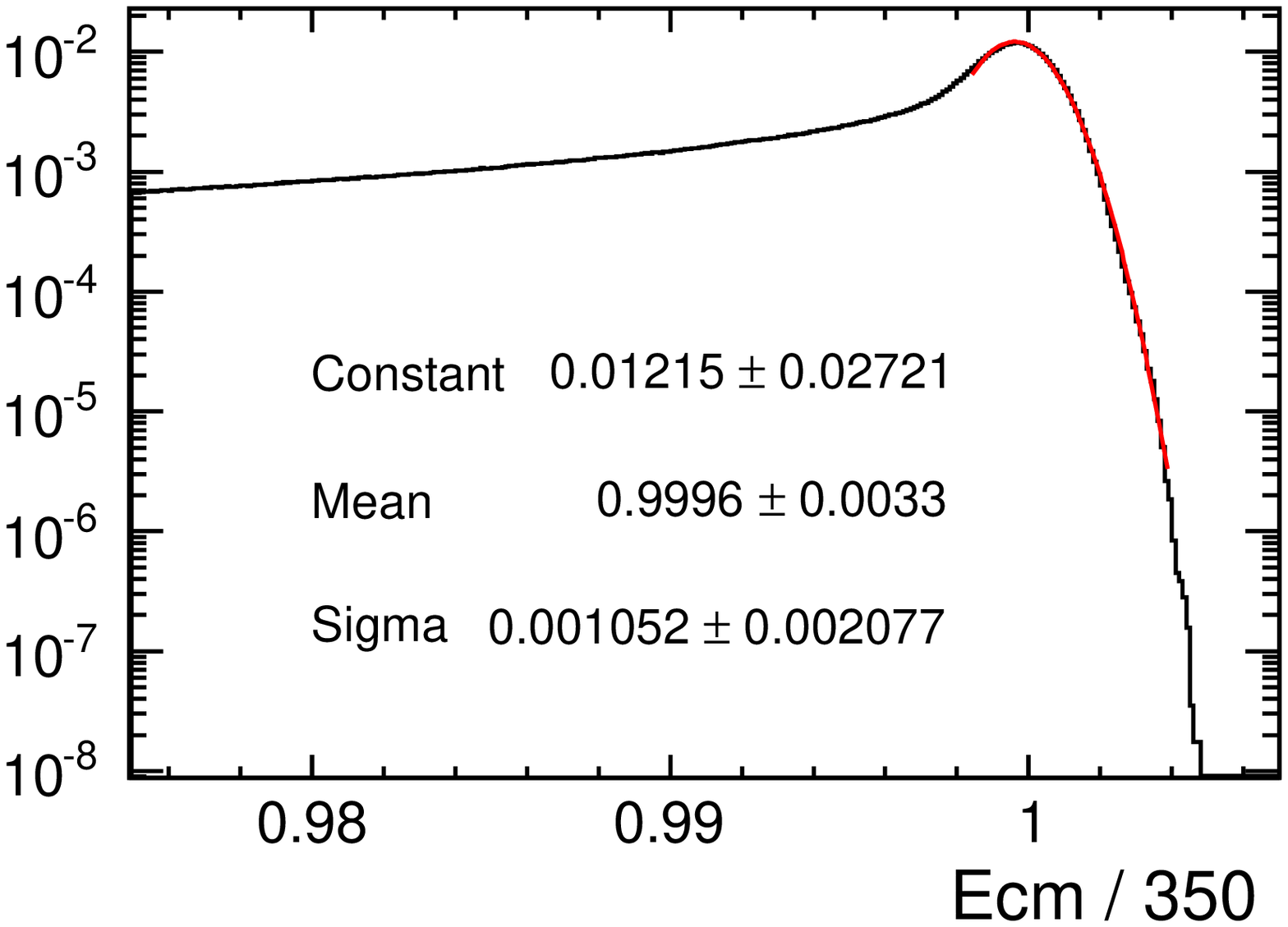} \label{fig:lumispe2}}
    \caption{Luminosity Spectrum at 350 GeV}
    \label{fig:lumispec}
  \end{center}
\end{figure}

The $\overline{\rm MS}$ mass is the most suitable mass scheme in most of the top physics.
In this study, we adopt the {\it Potential Subtracted mass} (PS mass)\cite{PSmass}, which is considered
to have the least correlation to the strong coupling constant ($\alpha_s$) in extracting the top mass, and then 
convert the PS mass to the $\overline{\rm MS}$ mass.
We perform a fit to the cross section as a function of E$_{\rm CM}$ floating the PS mass and the top width. 
Here we fix $\alpha_s(m_Z)$ to 0.12, expecting that it will be determined with much higher precision by the time the 
ILC starts. The top cross sections were calculated at NNLO in QCD. 
Figure \ref{fig:theocross} shows the calculated cross sections for various top masses and widths. 
The assumed integrated luminosity is the same as in the study of the top quark Yukawa coupling.
In extracting the mass and the width, we have to consider initial state radiation (ISR), beamstrahlung, 
and beam energy spread, which significantly affect the cross section curve.
Figure \ref{fig:lumispec} shows the luminosity spectrum of the ILC at E$_{\rm CM}$ = 350 GeV.
We use this spectrum for all the beam energies with energy scaling.
Since the theoretical calculation has been done assuming monochromatic energies,
we convolute the theoretical cross section with the beam spectrum as in the following equation.

\begin{equation}
  \sigma_{conv}(\sqrt{s}) = \int_0^1 {\cal L}(t)\ \sigma_{th}(t)dt    \nonumber
\end{equation}            
where $t = \sqrt{s'}/\sqrt{s}$ (where $\sqrt{s'}$ is the collision energy which is affected by beam effects and 
$\sqrt{s}$ is the nominal center of mass energy) $\sigma_{th}$ is the NNLO total cross section before convolution,
${\cal L}(t)$ is the normalized luminosity spectrum, and $\sigma_{conv}$ is the NNLO total cross section 
after convolution. The theoretical cross sections to be convoluted have been calculated at every 100 MeV 
from E$_{\rm CM}$ of 330 to 351 GeV.

\subsubsection {Fitting and Result}

\begin{figure}[tphb]
  \begin{center}
    \subfigure[Obtained cross section error at various top masses.]{\includegraphics[width=10cm]{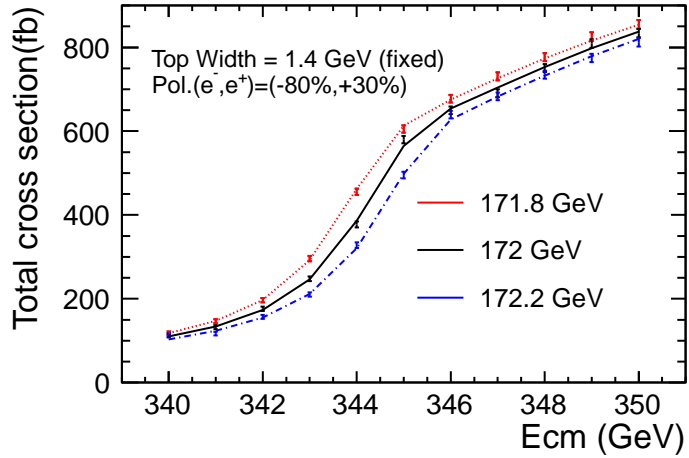} \label{fig:crossmt}}
    \subfigure[Obtained cross section error at various top widths.]{\includegraphics[width=10cm]{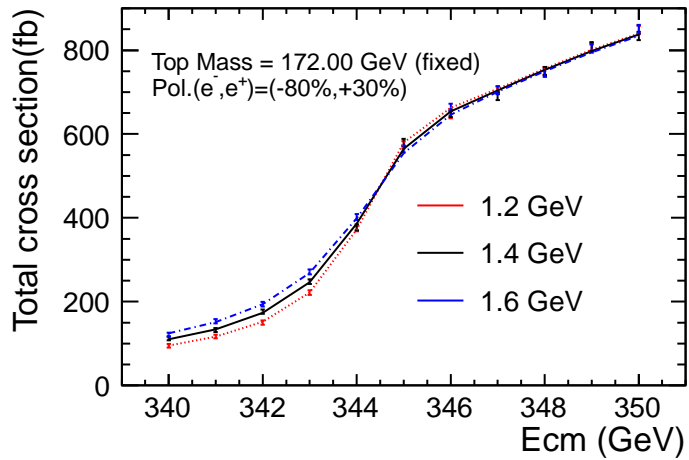} \label{fig:crosswt}}
    \subfigure[The correlation of top mass and width in the fitting result.]{\includegraphics[width=10cm]{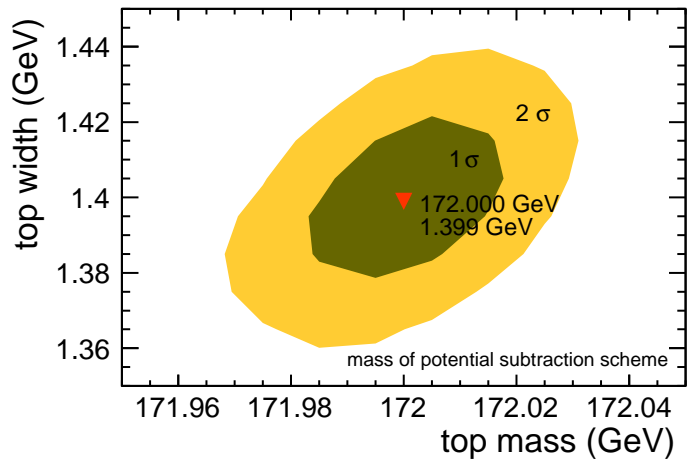} \label{fig:mtwt}}
    \caption{Fitting result}
    \label{fig:2dfit}
  \end{center}
\end{figure}

Using the convoluted NNLO cross sections we estimate statistical errors on the top mass and width 
with threshold scan at the ILC. Since the detector simulation studies have been done with the leading-order (LO) 
calculation, we assume that the signal selection efficiency and remaining background at NNLO are the same as in  
the LO analysis. The selection efficiency and background in the LO analysis are shown in 
Fig.~\ref{fig:CrossEff}, Table~\ref{tb:cutL}, and Table~\ref{tb:cutR}. The obtained number of signal events
which is scaled to the NNLO signal cross section and the number of background events after selection 
at each center-of-mass energy was randomized by Poisson distribution (toy-MC) and fitted to templates 
with two free parameters, namely the of top mass and its width. We produced the template samples of 
$m_t^{\rm PS} = 171 - 173$ GeV (at every 5 MeV in $m_t^{\rm PS} = 171.80-172.20$ GeV and every 10 MeV in other region)
and $\Gamma_t = 0.9-1.9$ GeV at every 10 MeV, resulting in over 24,000 templates. Linear interpolation was used 
for parameters between templates. Minuit2Minimizer\cite{rootmin} in ROOT was used for the minimization. 
10,000 toy-MC experiments have been performed. Table \ref{tb:psresult} shows the obtained mass and width errors and 
Figure \ref{fig:2dfit} shows the correlation of the two parameters as well as convoluted cross sections 
with measurement errors at several parameter values. We obtained 16 MeV statistical error for the top quark mass 
in the PS scheme and 21 MeV for the top width with the ``Left'' and ``Right'' results combined.  \\

\begin{table}[tphb] 
  \begin{center} 
    \begin{tabular}[bt]{|c|r|r|} \hline 

      & PS Mass (GeV) & Width (GeV)  \\ \hline 
      ``Left'' (110 fb$^{-1}$) & 172.000 $\pm$ 0.020 & 1.399 $\pm$ 0.026   \\ \hline 
      ``Right'' (110 fb$^{-1}$) & 172.000 $\pm$ 0.028 & 1.398 $\pm$ 0.038  \\ \hline 
      ``Left'' + ``Right'' (220 fb$^{-1}$) & 172.000 $\pm$ 0.016 &  1.399 $\pm$ 0.021  \\ \hline 
    \end{tabular}  
    \caption{Obtained PS mass and width with statistical errors, assuming 10 fb$^{-1}$ integrated luminosity 
      each for the 11 energy points from 340 to 350 GeV. The input PS mass and the width in the dataset are 172 and 1.4 GeV, respectively.}
    \label{tb:psresult}
  \end{center}
\end{table} 

The conversion from the PS mass to the $\overline{\rm MS}$ mass can be written as
\begin{equation}
  m_t^{\overline{\rm MS}}  \ \sim \ m_t^{\rm PS} \ - \ \cfrac{4}{3\pi}\ (m_t^{\rm PS} -20)\ \alpha_s +\ \ldots  .  \nonumber
\end{equation}
Using $\alpha_s$ of PDG value, we obtain $m_t^{\overline{\rm MS}}$ = 163.80 $\pm$ 0.016(stat) .
   
\section{Summary}\label{sec:summary}
Measuring properties of the top quark is quite important to test the SM and search for physics beyond the SM. 
By the threshold scan, the top properties such as its mass, width, and Yukawa coupling, will be able to be measured 
accurately at the ILC.   \\
We have estimated the statistical errors on the top mass, width, and Yukawa coupling 
using the ILD simulation framework. In our study, 11 energy points (between 340 and 350 GeV) and 
two beam polarization combinations (P($e^+,\ e^-$) = ($\pm$0.3, $\mp$0.8)) with 10 fb$^{-1}$ each, 220 fb$^{-1}$ 
in total, are used for the threshold scan. Only 6-Jet final state, $t\overline{t} \rightarrow bWbW \rightarrow bqqbqq$, 
was considered as the signal in this study. For the top quark Yukawa coupling, 4.2\% statistical error was obtained. 
For the top mass and width, NNLO total cross section was used to scale the LO analysis. 
We obtained $\delta m_t$ = 16 MeV for the potential subtracted mass and $\delta \Gamma_t$ = 21 MeV for the top width.

Our results were compared with the previous study\cite{Martinez:2002st} and the study at the CLIC\cite{clic}. 
According to the previous study, the prediction of determination of the top quark mass, with an experimental 
accuracy better than 30~MeV and of width, with an accuracy at the 2\% level, is quit robust. 
Therefore $\delta m_t$ = 14~MeV and $\delta \Gamma_t$ = 18~MeV which were obtained when our results were scaled 
from 220 to 300~fb$^{-1}$ were consistent. 
Since the study at the CLIC, which used the luminosity spectrum of the ILC and CLIC\_ILC detector, 
estimated $\delta m_t$ = 27~MeV, $\delta m_t$ = 24~MeV which was our result scaled from 220 to 100~fb$^{-1}$ was also consistent.

We plan to add 4-Jet final states, which has similar branching ratio to 6-Jet, to improve the sensitivities. 
Since sensitivities can be improved by optimizing the strategy of the threshold scan, we also plan 
to study several running scenarios.

\vspace{1.0cm}
\hspace{0.2cm} {\bf Acknowledgments}
\vspace{0.5cm}

The authors would like to thank all the members of the ILC physics subgroup \cite{subgroup} for 
useful discussions on this work and those of the ILD software and optimization group, who maintain 
the software and Monte Carlo samples used in this work. 
This work is supported in part by the Creative Scientific Research Grant No.~18GS0202 of the Japan Society 
for Promotions of Science (JSPS), the JSPS Grant-in-Aid for Science Research No.~22244031,
and the JSPS Specially Promoted Research No.~23000002.

\end{document}